\documentclass[aps,prl,twocolumn,groupedaddress]{revtex4-2}

\usepackage{graphicx}
\usepackage{subcaption}
\usepackage{amsmath,amssymb}
\usepackage{bm}
\usepackage{booktabs}
\usepackage{multirow}
\usepackage{amsthm}
\usepackage{caption}
\theoremstyle{definition}   
\newtheorem{hypothesis}{Hypothesis}
\newtheorem{remark}{Remark}

\begin{document}

\title{Universal Extremum Seeking Mechanism for Lift Variation in Soaring Birds Flight: \\ A New Paradigm in Computational Physics and Biology}

\thanks{eisash@ucmail.uc.edu}
\thanks{This research was developed with funding from the Defense Advanced Research Projects Agency (DARPA). The views, opinions and/or findings expressed are those of the author and should not be interpreted as representing the official views or policies of the Department of Defense or the U.S. Government. Distribution Statement “A” (Approved for Public Release, Distribution Unlimited).}

\author{Simone Martini, Dipesh Kunwar, Sameh A. Eisa}

\affiliation{Department of Aerospace Engineering and Engineering Mechanics, University of Cincinnati, Cincinnati, OH 45221 USA}

\begin{abstract}
In this letter, we reveal a universal, very simple extremum seeking natural feedback law and mechanism that governs, adapts, and generates in real-time, optimized lift variations for successful energy gain flight in presence of wind shear. The introduced law/mechanism, which is computationally minimal and needs only sensory information of the wind or local energy rate (i.e., model-free and data-driven) is able to characterize and replicate dynamic soaring optimized flight physics of windward climb in real-time for a variety of soaring birds species, namely wandering albatross, black-browed albatross and grey-headed albatross. We confirm the effectiveness of this new simple, real-time law by successful comparisons with sophisticated non-real-time optimal control solver and reported biological data. Our results establish the proposed mechanism as a new paradigm in soaring flight physics. That is, our results substantially advance the computational physics/biology aspects of the problem while providing a biologically plausible theory for avian soaring behavior. 
\end{abstract}

\maketitle
\begin{figure}[ht]                       
    \centering	\includegraphics[width=0.40\textwidth]{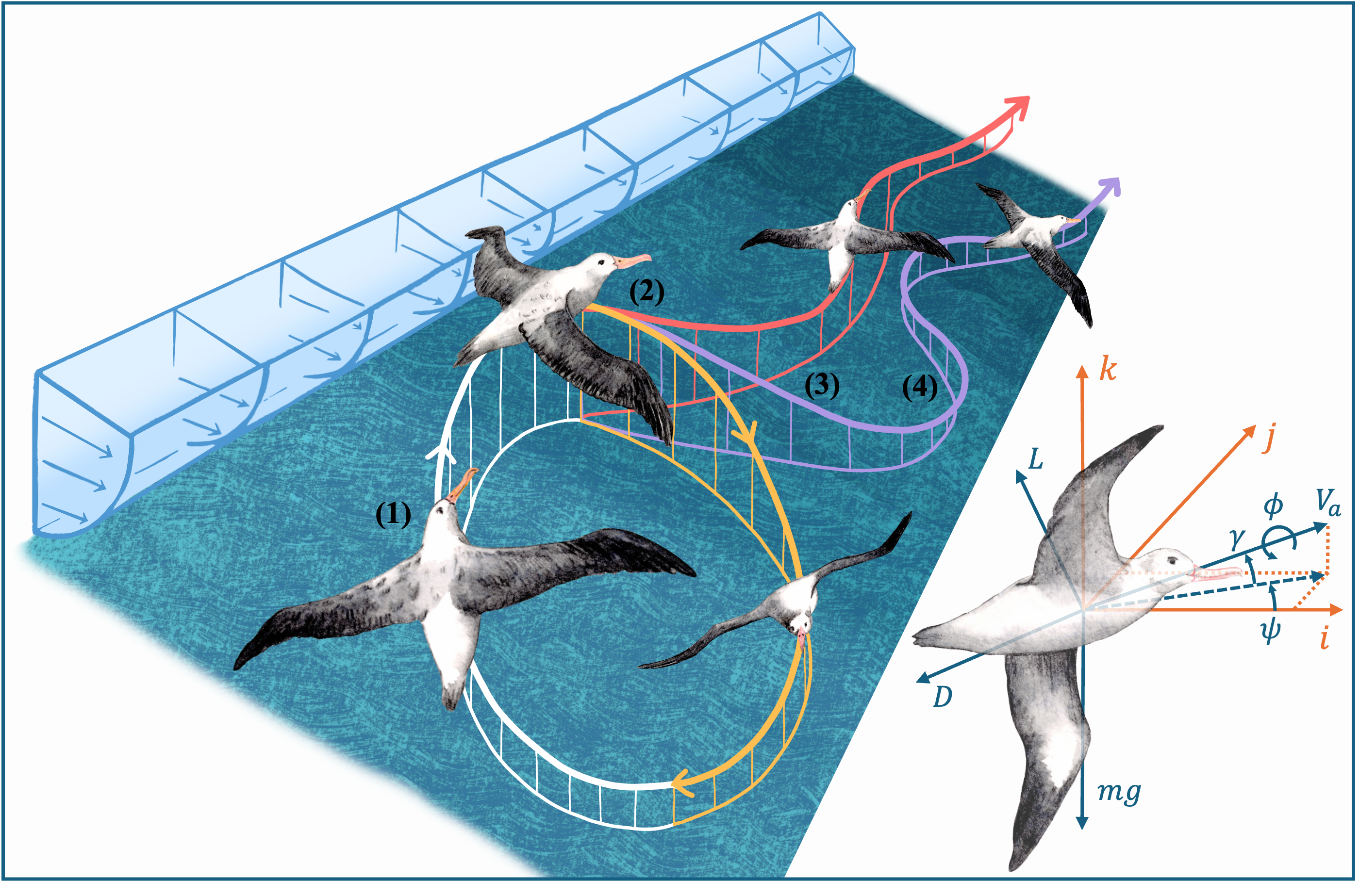}
    {\\ \footnotesize Artwork credits: Claudia Acutis.}
    \caption{\footnotesize{The dynamic soaring maneuver (DS) of a wandering albatross, characterized by four distinct flight phases:  (1) windward/upwind climb, (2) high altitude turn, (3) leeward/downwind descent, and (4) low altitude turn. The maneuver is illustrated in presence of a logarithmic wind profile above the surface of the ocean. The upwind climb phase (white trajectory), allows for initiating the wind energy extraction and enables the DS cycle to take several shapes according to the overall bird's objective and environmental conditions: (red) task-oriented/traveling DS, (purple) cyclic/periodic DS, and (yellow) loitering DS (common during foraging). The zoomed-in section shows the adopted system of reference.}}
    \label{fig:dynamicsoaring}
\end{figure}

 The optimized soaring flight physics, often called dynamic soaring (DS) is an extremely energy-efficient maneuver performed by soaring birds to extract energy from wind and drastically extend their travel capabilities during foraging and migratory expeditions. Extraordinary statistics about this phenomenon can be found in the review article \cite{mir2018review} summary of which is given in this video \cite{MDCL2022DynamicSoaring}. Ideal versions of DS maneuvers are generally characterized by four phases represented in Fig. \ref{fig:dynamicsoaring}: windward/upwind climb, high altitude turn, leeward/downwind descent, and low altitude turn.
Although observation of this phenomenon has been reported as early as the 16th century by Leonardo Da Vinci \cite{richardson2019leonardo}, its first physical characterization was proposed by Lord Rayleigh in 1883 \cite{rayleigh1883soaring}. Since then, considerable scientific effort has been put into better unpacking of such efficient evolutionary skill. Although DS was initially seen as a purely periodic maneuver \cite{rayleigh1883soaring}, recent real-world observations including GPS tracking data \cite{richardson2022observations,weimerskirch2000fast,richardson2018flight}, \cite[Fig. 4]{sachs2013experimental}, \cite[Fig. 1]{mir2018review}, have suggested that DS is the result of optimally adapting the flight trajectory and aerodynamic characteristics to environmental conditions, mainly by controlling the attitude and lift coefficient configuration through pitching and/or morphing of the wing shape, and using soaring birds innate sensory ability to close the control feedback loop \cite{pennycuick2008modelling, MIR201817, WANG2024317}. It is, in fact, believed that soaring birds guide their forage through olfactory search \cite{doi:10.1073/pnas.0709047105} and some speculate that their nostrils have evolved to perceive wind variations \cite{pennycuick2002gust,mangold1946nase}. Hence, a considerable research effort has been devolved to solving DS as an optimal control problem \cite{sachs2005minimum,mir2018review}. Given the nonlinear and multivariate nature of bird flight dynamics and wind dynamics, only linear-wind-profile simplification of DS can be analytically solved \cite{lawrance2009guidance}, while nonlinear-wind-profile DS solutions can be achieved with numerical optimal control solver such as GPOPS-II \cite{gpops2}, BOUNDSCO, and TOMP \cite{sachs1991optimal,zhao2004optimal,mir2018review}. Although the latter approach is able to compute/predict energy efficient DS trajectories taken or to-be-taken by soaring birds, it presents some key criticalities: i) reliance on a mathematical description of the underlying physics (i.e. model-based); ii) high computational demand which makes this approach non-real-time implementable; iii) strong dependence on boundary conditions (which might not always be known); and iv) open loop approach since the control input trajectory is generated once and cannot be realistically updated throughout the maneuver. These criticalities do not only pose an implementation-related concern but most importantly highlight a biological incompatibility problem by not being real-time adaptable. This invokes the question: what mechanism would allow soaring birds to execute an optimized control response with real-time adaptation to successfully perform DS as observed in nature? 

With this rationale, in recent findings \cite{eisa2023analyzing}, extremum seeking control (ESC) was, for the first time, employed as model-free, real-time, and computationally-inexpensive optimization strategy to characterize the DS maneuver of soaring birds such as the wandering albatross. Said characteristics of the ESC highlight its biological and physical plausibility as opposed to optimal control solvers which are generally model-based and operate in a non-real-time setting. While the aforementioned literature pioneered the use of ESC for DS, it does so in a ``basic mode setting" for which the lift coefficient $C_L$ is fixed and the ESC strategy is applied to adapt the roll/bank angle $\phi$ \textit{only} to replicate a single ideal DS cycle with energy-neutrality. Conversely, in this letter we propose, for the first time, a universal ESC adaptation law for $C_L$, responsible for controlling lift variation, to characterize and replicate/predict, general DS maneuvers such as those provided in Fig. \ref{fig:dynamicsoaring}. We confirm the effectiveness of this new simple, real-time law by successful comparisons with sophisticated non-real-time optimal control solver and real-world biological data. 

Throughout this letter, we establish our proposed universal ESC law by addressing, validating and providing evidences to the following hypotheses:
\begin{hypothesis}\label{hp:2} \textit{(the proposed law/mechanism successfully characterize the optimized flight physics of soaring)}. 
The proposed ESC law is a universal, very simple natural feedback law and
mechanism that governs, adapts, and generates in real-time, optimized lift variations for successful
energy gain flight in presence of wind shear. 
\end{hypothesis}

\begin{hypothesis}\label{hp:3}
\textit{(the proposed law/mechanism is biologically plausible).}
The proposed ESC mechanism adapts in real-time, the lift variations based on simple, model-free, sensory-based and computationally-basic feedback law (online, simple, data-driven learning mechanism).   
\end{hypothesis}

\begin{hypothesis}\label{hp:1}\textit{(the proposed law/mechanism leads to previously impossible advances in computational biology and computational physics of DS)}.
The proposed ESC mechanism is an effective representation of the biophysics of DS that only needs basic models to simulate/predict optimized soaring in real-time with basic computational demand, which leads to breakthroughs in the computational cost and complexity of solving DS, allowing to generate millions of accurate soaring flight trajectories in only few hours with regular computers.
\end{hypothesis}

\begin{figure*}[htb!] 
    \centering
    \includegraphics[width=0.95\textwidth]{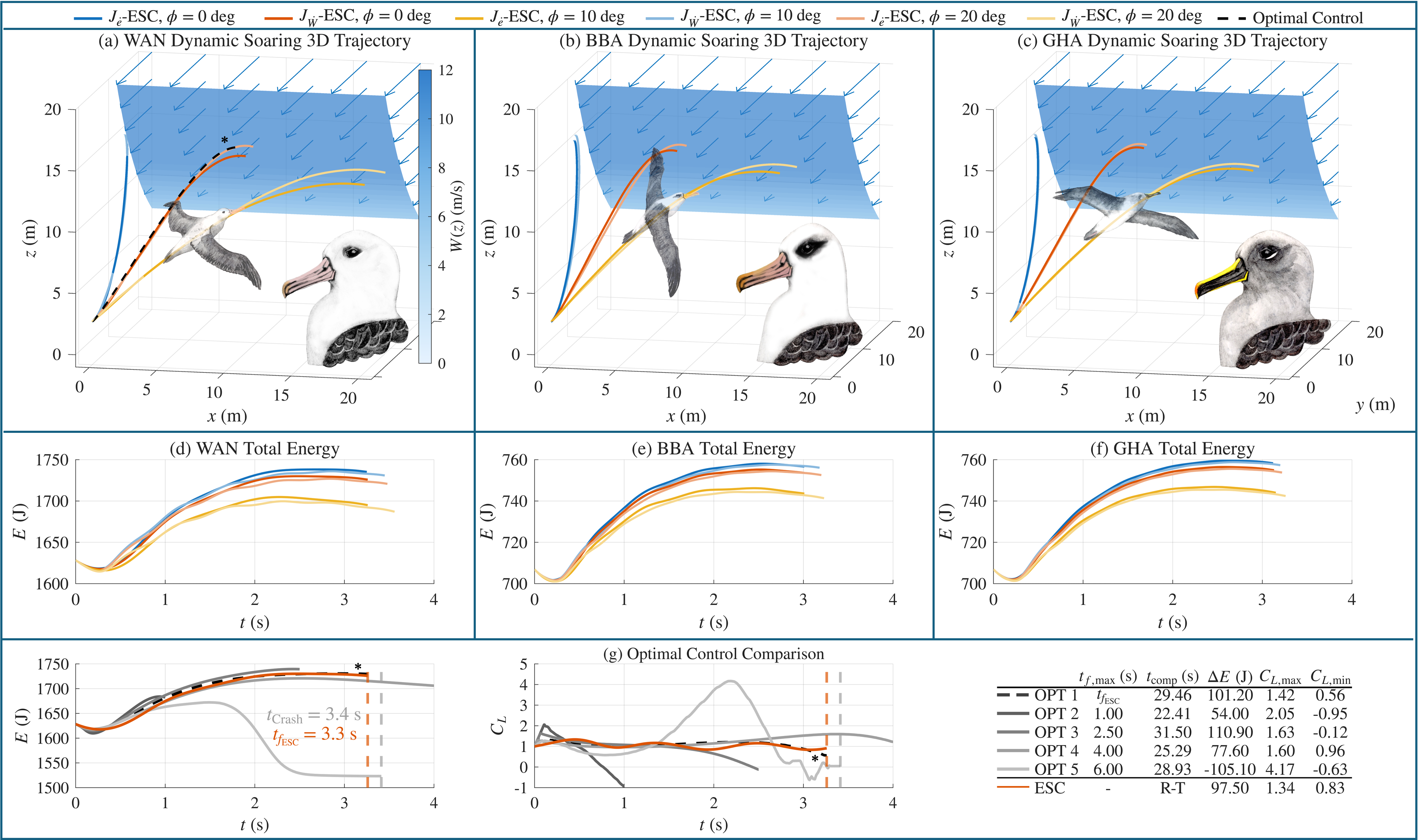}  
    {\footnotesize \\ Albatross illustrations credits: Claudia Acutis.}
    \caption{(a), (b) and (c) show the ESC system trajectories at different fixed roll angle $\phi$ of three albatrosses species WAN, BBA and GHA, respectively. Darker shades curves refer to trajectories generated based on the measurement of $J_{\dot{e}}$ (local energy gain) while lighter shade curves refers to the ones generated by the measurement of $J_{\dot{W}}$ (wind shear). The black dotted line in (a) shows a WAN
    trajectory with fixed $\phi = 10$ deg  
    generated by the optimal control software GPOPS-II by setting the final time bound $t_{f_{\text{max}}}$ equal to the final time resulting from the ESC simulation (with $J_{\dot{e}}$ and fixed $\phi = 10$ deg), $t_{f_\text{ESC}} = 3.26\, \text{s}$. (d), (e) and (f) display the total energy $E$ of the trajectories in (a), (b) and (c), respectively. ESC with $J_{\dot{e}}$ and $J_{\dot{W}}$ measurements yield almost identical energy gains $\Delta E$, ranging from $58 \leq\Delta E_{\text{WAN}} \leq 107 $ J for  WAN in (d), $34 \leq\Delta E_{\text{BBA}} \leq 50 $ J for  BBA in (e), and $35 \leq\Delta E_{\text{GHA}} \leq 51 $ J for  GHA in (f). Visual inspection of (a), (b) and (c) in relation with  (d), (e) and (f), clearly highlights the tradeoff of 
    $\Delta E$ for across-wind traveled distance.
    (g) Shows $E$ and $C_L$ of 10 degrees roll-fixed DS trajectories of WAN computed by GPOPS-II by setting several values of the maximum time bound $t_{f_{\text{max}}}$. The asterisk highlights that the dotted line shown in (g) is the same optimal trajectory displayed in (a). The legend reports the respective values of $t_{f_{\text{max}}}$ which a user of the optimal control software might chose, the computational time $t_{{\text{comp}}}$ of each optimal control generated trajectory, the energy gain $\Delta E$, and maximum and minimum values of $C_L$. When the optimal control solver has a too short final bound  $t_{f_{\text{max}}} = 1$, the obtained $\Delta E$ is lower than the ESC one, when it is set to high $t_{f_{\text{max}}}=6$s, the optimal solver breaks and returns a negative net-energy results with $t_f = t_{f_{\text{Crash}}} = 3.4\, \text{s}$ (light gray vertical dotted line). The optimal control curve with $t_{f_{\text{max}}}=2$ yields a higher $\Delta E$ but using unfeasible values of $C_{L_\text{max}}$ and $C_{L_\text{min}}$. When $t_{f_{\text{max}}}$ is set equal to $t_{f_\text{ESC}}$ (red vertical dotted line), the energy gain of the optimal control solver and the ESC trajectories is practically the same. Lastly, the computational time $t_{{\text{comp}}}$ to generate the optimal control trajectories ranges from about 20 s to 30 s, while the ESC counterpart is computed in real-time (R-T) throughout the DS maneuver. One can conclude that the proposed ESC law/mechanism achieves optimal performance comparable to the optimal control solver while presenting two crucial benefits: (i) there is no need to provide a guess for the bounds to achieve a physical solution and (ii) computational times are reduced by more than three orders of magnitude.}
    \label{fig:2}
\end{figure*}

\textit{Universal ESC law for $C_L$ Adaptation}--To model the dynamics of soaring birds, we consider the standard \cite{eisa2023analyzing} thrust-less point mass DS model in the inertial reference frame including wind relative kinematics, expressed in the control-affine form
\begin{equation}\label{eqn:controlAffine}
    \dot{\bm{x}}=\bm{f}(\bm{x})+\sum_i^m \bm{b}_i(\bm{x}) u_i(t),
\end{equation}
where state $\bm{x}\in\mathbb{R}^n$, drift dynamics $\bm{f}(\bm{x})$, control vector fields $\bm{b}_i(\bm{x})$, and control inputs $u_i(t)$ are expressed as 
\begin{align}\label{eqn:proposedStructure_states}
    \bm{x} &= \begin{bmatrix}
    x &
    y&
    z&
    V_a&
    \gamma&
    \psi&
    \phi &
    C_L
    \end{bmatrix}^{\top},\\
    \bm{f}(\bm{x}) &= \begin{bmatrix}
    V_a \cos\gamma \cos\psi\\
    V_a \cos\gamma \sin\psi-W\\
    V_a \sin\gamma\\
   \frac{1}{m}( -D -mg\sin\gamma+m\dot{W} \cos\gamma \sin\psi)\\
    \frac{1}{mV} (L\cos \phi - mg \cos \gamma - m \dot{W} \sin \gamma \sin \psi) \\
    \frac{1}{mVcos \gamma}(L \sin \phi + m \dot{W} \cos \psi) \\
    0 \\
    0 \\
    \end{bmatrix},\label{eq:drift}\\
        \bm{b}_1 &= \begin{bmatrix}
    0 & 0 & 0 & 0 & 0 & 0 & 1 & 0 
    \end{bmatrix}^{\top},
        \bm{b}_2 = \begin{bmatrix}
    0 & 0 & 0 & 0 & 0 & 0 & 0 & 1 
    \end{bmatrix}^{\top},\\
    \dot \phi &= u_1, \qquad \dot C_L = u_2.
\end{align}
where $x,y$ and $z$ are the inertial spatial coordinates, $V_a$ is the body airspeed, $\gamma$ is the flight path angle, $\psi$ is the heading angle, $W$ is the wind speed, $m$ is the body mass, $g$ is the gravitational acceleration, and $D$ and $L$ are the drag and lift, respectively, which are functions of $C_L$ as
\begin{equation}\label{eqn:aero_forces}
L = \tfrac{1}{2}\rho V_a^2 S C_L, \quad D = \tfrac{1}{2}\rho V_a^2 S  \underbrace{(C_{D0} + \frac{1}{\pi \textit{AR} e} C_L^2)}_{C_D}
\end{equation}
with $\rho$ being the air density, $S$ the wing area, $C_{D0}$ the zero-lift drag coefficient, $\textit{AR}$ the wing aspect ratio, $e$ is the Oswald efficiency factor, and $C_D$ the drag coefficient.
DS maneuvers evolve through the wind shear by extracting energy from the wind with speed and gradient, are generally estimated through a logarithmic model \cite{pennycuick1982flight, mir2018review}: 
\begin{equation}\label{eqn:wind_log}
    W(z)=V_{W_{\text{ref}}}\frac{\ln(z/h_0)}{\ln(z_{\text{ref}}/h_0)}, \quad \dot W(z,\dot z) = V_{W_{\text{ref}}}\frac{\dot z}{z\ln(z_{\text{ref}}/h_0)}
\end{equation}
where $V_{W_{\text{ref}}}$ is the measured reference wind speed at a reference altitude $z_{\text{ref}}$, and $h_0$ is the lower admissible altitude limit and corresponds to the roughness factor of the terrain.
The model \eqref{eqn:controlAffine} is widely adopted in the literature to  generate DS trajectories using model-based optimal numerical solvers \cite{sachs2005minimum,sukumar2013sailplanes,liu2017bio}. Although we employ \eqref{eqn:controlAffine} for numerical simulations, the ESC mechanism is not based on the knowledge of the underlying dynamics but solely on the measurement of an objective function. In fact, ESC is a model-free adaptive control strategy which steers the dynamics toward the extremum (maximum/minimum) of an objective function -- see \cite{scheinker2024100} for a comprehensive review. A general ESC law for the $i$-th input channel is expressed as
\begin{equation}\label{eq:ESC_law}
    u_i = b_{i,1}(J(\bm{x}))\sqrt{\omega}a_{i,1}u_{i,1}(t) + b_{i,2}(J(\bm{x}))\sqrt{\omega}a_{i,2}u_{i,2}(t),
\end{equation}
where $b_{i,1}(\cdot)$ and $b_{i,2}(\cdot)$ are functionals of the objective function to be optimized $J(\bm{x})$, $a_{i,1}$ and $a_{i,2}$ are control parameters, and $u_{i,1}(t), u_{i,2}(t)$ are periodic and zero-average dither signals with angular frequency $\omega$.

As shown in Fig. \ref{fig:dynamicsoaring}, real-world observations indicate that DS maneuvers are not characterized by fixed periodic movements but can result in a variety of paths, leaving the soaring bird with a margin of adaptation based on environmental conditions and 
overall mission goal. 
Nevertheless, characteristic maneuver that enables all DS paths is the \textit{windward climb} during which, the energy \textit{extraction from the wind} is initiated and transferred to the bird \cite{sachs2013experimental}. From our observation (later showcased in Fig. \ref{fig:2}), the energy gained in the upwind climb phase, not only offsets the drag dissipation, but can also provide the soaring bird with a surplus 
(in form of kinetic or potential energy) which can possibly be used to perform less efficient but more task-oriented maneuvers. Analyzing a \textit{pure} windward climb 
(i.e. maintaining constant heading and fixed roll angle $\phi = 0$), 
it becomes clear that this maneuver is primarily affected by behavior of the 
lift coefficient $C_L$. 
Albatrosses' ability of adapting their $C_L$ by pitching and/or wing-shape morphing has been studied and characterized in \cite{MIR201817} and \cite{WANG2024317}, showing that, during DS maneuvers, albatrosses $C_L$ can range from values of 0 to 1.5 \cite[Fig. 8]{WANG2024317}. Hence, in this letter we propose an ESC mechanism to adapt lift variation and enable optimal energy gain during upwind climb. To this end, 
set
\begin{equation}
    b_{1,1} = 0, \quad b_{1,2} = 0, \quad b_{2,1} = J(\bm{x}), \quad b_{2,2} = 1,
\end{equation}
which makes it so that $C_L$ is adapted to reach the extremum of the objective function $J(x)$ while the dynamics of $\phi$ are not perturbed by the ESC dither.

For the choice of $J(\bm{x})$ we consider the system's specific energy (total energy $E$ per unit weight) as in \cite{eisa2023analyzing}:
\begin{equation}\label{eqn:energy}
    e = \frac{E}{mg} = z + \frac{V_a^2}{2g},
\end{equation}
differentiating with respect to time yields
\begin{equation}\label{eqn:energyGain}
\begin{split}
    \dot{e} = \dot{z}+\frac{V\dot{V}}{g} = -\frac{DV_a}{mg}+\frac{V_a\dot{W}\cos \gamma \sin \psi}{g}.
\end{split}
\end{equation}
The two terms in \eqref{eqn:energyGain} represents the energy depletion due to drag resistance and energy harvested from the wind shear, respectively. We propose the following two objective functions to be maximized
\begin{equation}\label{eq:J}
    J_{\dot{e}}(\bm{x}) = \dot{e}, \quad J_{\dot{W}}(\bm{x}) = \dot{W}.
\end{equation}
The maximization of $J_{\dot{W}}$ aims at generating a $C_L$ adaptation such that the birds travel through the highest value of wind shear \textit{needing only to sense it}, hence increasing the wind energy extraction. On the other hand, $J_{\dot{e}}$ aims at maximizing total energy gain inducing a tradeoff between wind energy extraction and drag resistance. In essence, the choice of these objective functions reflects the maximization of direct physical quantities. Note that, only local measurements of the energy state and the environmental conditions are sensed while measurement of aerodynamic drag is not needed.
\begin{remark}\label{rm:1}
    The ESC law \eqref{eq:ESC_law} applied to the  control affine soaring bird model \eqref{eqn:controlAffine} is extremely simple to implement, simulate, and to compute.
\end{remark}

\textit{Evaluating the Proposed ESC Law via Simulations and Comparison with Optimal Control Solvers}--
To test the $C_L$ ESC mechanism, we select three albatross species' parameters from \cite{pennycuick1982flight}, which are known in literature as some of the most efficient soaring birds, namely wandering albatross (WAN), black-browed albatross (BBA) and gray headed albatross (GHA). Although the rationale for modulating $C_L$ comes from the pure upwind climb, \cite{sachs2013experimental} highlights that most of the energy is gathered during the \textit{upper curve} (high altitude turn). Hence, by selecting different fixed values of $\phi$ we can simulate several \textit{non-pure} upwind climb maneuvers which merge/connect the windward climb with the high altitude turn as shown in Fig. \ref{fig:2} (a-c). For PC's software and hardware setup the reader should reference the supplemental material (file1.pdf).
From Fig. \ref{fig:2} (d-f), we can verify that the ESC law allows for net energy gain during the upwind climb maneuver of all three species and, the more banked is the roll angle, the more across-wind travel is induced at the expense of less amount of energy being gathered. 
We also highlight that when we use optimal control solver such as GPOPS II for DS, one might not know a priori all the necessary boundary conditions, hence one would be forced to feed a guess to the numerical optimization solver, which might influence the optimization itself, or even worse, render it unfeasible (leading to instability of the numerical solver). Fig. \ref{fig:2} (g) illustrates, for instance, how a researcher might approach the DS problem using GPOPS-II optimal software and how, out of the feasible solutions that the optimal solver provides, the one resulting in the highest energy gain is the one that set the maximum final time bounds equal the final climb time achieved by the ESC system.
Additionally, we remark that the optimal solver works in a open loop fashion, after the model-based computation of a solution, ranging around 20 s to 30 s, the control input would have to be applied to the system without any possibility of feedback adaptation. On the other hand, the ESC is directly applied to the system dynamics, and adapt in real-time throughout the climb (with the whole simulation taking roughly 0.015 s).
This observation exposes a major drawback in using numerical optimal solver for DS since the terminal bounds are themselves an optimal result influenced by the external environment (i.e. they require knowing the final answer to compute the solution).
Conversely, because of its real-time optimal adaption capabilities, achieved by a simple and low cost mechanical process, we propose the ESC as a universal mechanism for lift variation in DS. 
\begin{figure*}[htb]  
    \centering
    \includegraphics[width=0.95\textwidth]{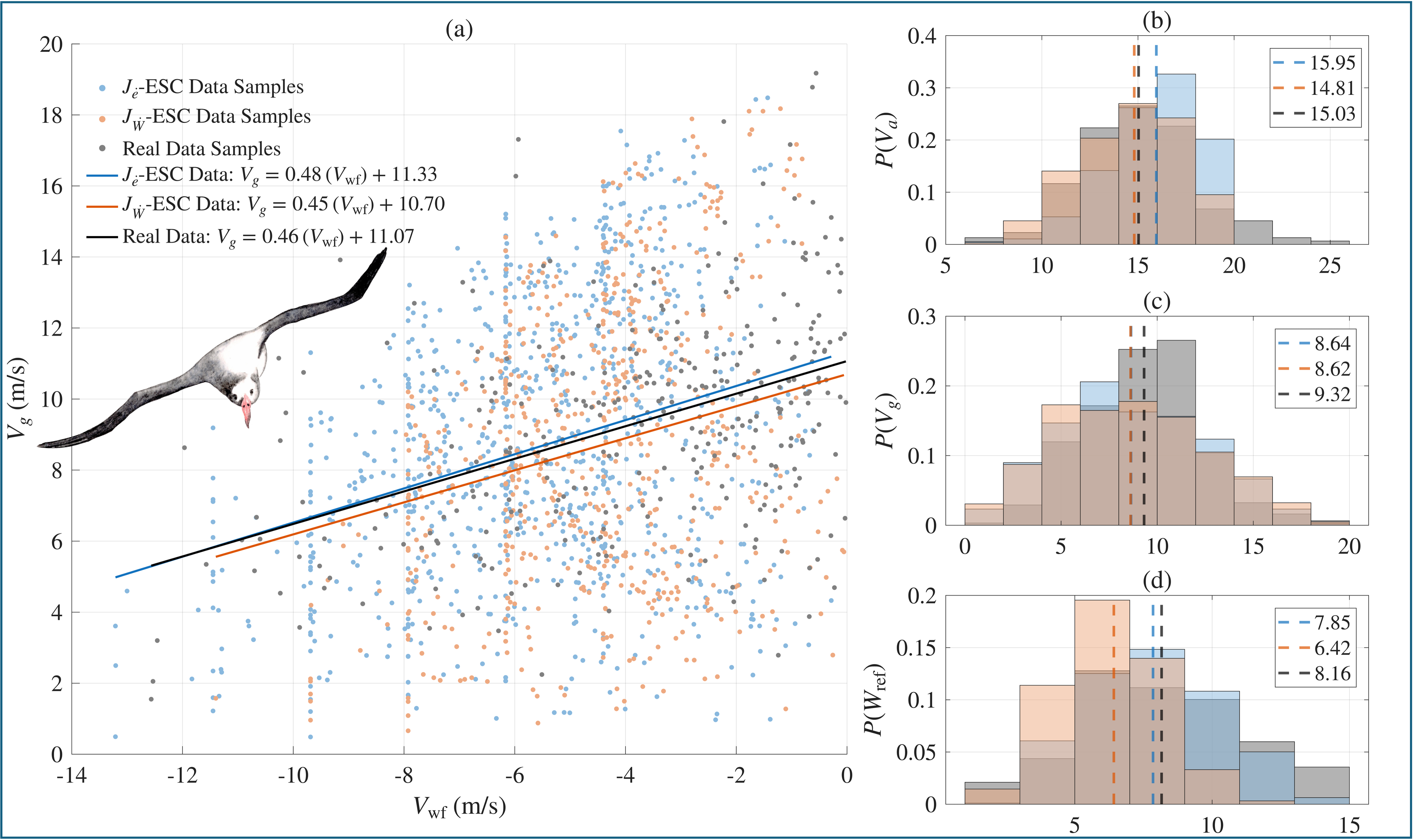}  
        {\footnotesize \\ Albatross illustration credits: Claudia Acutis.}
\caption{Comparison of ESC DS dataset with biological data from \cite{richardson2018flight}. On the left, (a) shows the scatter plot
and 
regression lines of biological data and the ESC dataset with measurement function $J_{\dot e}$ (local energy gain) and $J_{\dot W}$ (wind shear). The ESC regression lines differ from the biological data counterpart by less the 5\% in terms of slope and intercept percentage error, which shows that the ESC dataset successfully captures the DS macroscopic physical relation of groundspeed $V_g$ with respect to wind in the direction of flight $V_{\text{wf}}$. On the right, sub-figures (b), (c) and (d), present the overlap of probability density histograms of airspeed $V_a$, $V_g$, and reference wind $W_{\text{ref}}$ data samples, respectively. The histograms are associated with a dotted line displaying the mean value. Histograms and mean line colors reflect the color of the samples in the scatter plot's legend. The ESC DS dataset sample distribution has generally low $L_2$ and Wasserstein distance $D_W$ (file1.pdf) with respect to the biological data from \cite{richardson2018flight}. The highest $D_W$ of $1.955$ m/s, registered in the $W_{\text{ref}}$ distribution of the ESC dataset with measurement function $J_{\dot W}$, is less than the dataset discretizations value of 2 m/s. 
The overall accordance of the sample distribution shows that the ESC DS dataset captures the preference/requirements of the flight conditions in which the albatrosses fly. Lastly, from strong agreement with the biological data from \cite{richardson2018flight}, one can conclude that the generated ESC DS dataset replicates the macroscopic physical and biological characteristics of DS.}
    \label{fig:fig3}
\end{figure*}
\begin{remark}\label{rm:2}
The DS windward climb is an extremely fast flying maneuver that takes very few seconds. Thus, cannot tolerate failure in response or adaptation as this can have substantial consequences on soaring birds flight success and safety. The proposed ESC mechanism is the first physical formulation which can successfully model the real-time $C_L$ evolution and adaptation required in a DS maneuver. The law is general since it accounts for a broad range of upward climb conditions under different roll angels and with different set of birds parameters.
\end{remark}
\begin{remark}\label{rm:3}
The $C_L$ ESC adaptation law requires only local measurements (sensation) of velocity and accelerations ($J_{\dot e}$) or wind variations ($J_{\dot W}$), which makes the produced $C_L$ physically feasible (through pitching and/or wing morphing) similar to what is observed biologically. In fact, Figure \ref{fig:2} (g) shows how the dither signals induce, for the $C_L$ adaptation, an optimal search which resembles a ``trial and error adaptation" behavior that fits biological learning processes \cite{skinner2019behavior,pearce2013animal}.
\end{remark}
\begin{remark}\label{rm:4}
     Even with basic and simple configuration that uses point-mass dynamic model, logarithmic wind profile, analytic drag formulation without any computational fluid dynamics (CFD), and accurate guess of the boundary condition, sophisticated optimal control solvers such as GPOPS II need thousands of times more computational time and power than the proposed ESC law.
\end{remark}

\textit{Accordance with Biological Data}--
We claim that a soaring bird performing DS can be modeled as an Extremum Seeking Control System (ESC), hence, we set out to model and simulate DS maneuvers and compare the resulting data with real biological data from literature \cite{richardson2018flight} and \cite{pennycuick1982flight}. In \cite{richardson2018flight}, the authors analyze groundspeed speed $V_g$ of WAN with respect to wind and publicly share the WAN data presented in \cite{wakefield2009wind}, which were gathered with up-to-date GPS technologies attached directly on the body of several albatrosses. On the other hand \cite{pennycuick1982flight} is a foundational study in the literature for which the author analyzed several soaring birds flight data, gathered with an ornithodolite \cite{pennycuick1982ornithodolite}, covering nine species, out of which, WAN, BBA, and GHA, consistently perform DS maneuvers. 
The reader should reference the supplemental material (file1.pdf) for simulation parameters and details.
The reason for comparing our ESC mechanism with two literature works is to provide evidence of a strong accordance of our generated dataset to real data. However, considering the data gathering technological advancements and the public availability of the dataset, we will assume a higher degree of reliability for the data in \cite{richardson2018flight}. Hence, our findings will be summed from the comparison with \cite{richardson2018flight} and double checked by the comparison with \cite{pennycuick1982flight}.

We proceed by creating the ESC DS dataset with the following rationale: (i) Identify the most relevant variable and initial conditions in DS climb, namely $C_{L_0}, V_{a_0}, z_{0}, \phi_{0}, \psi_{0}, V_{W_{\text{ref}}}$; (ii) For said initial conditions and variables, from literature works, identify plausible bounds and select a set of $n$ values within these bounds with $n = 1+(\text{max} - \text{min})/d$, where $d$ is the selected discretization; (iii) Simulate all possible combinations and extract the trajectories resulting in net energy gain; (iv) For each complete trajectory, select one random instant (through uniform sampling) as a data point; and (v) compare the obtained data points to biological data of literature.
The reader should reference the supplemental material (file1.pdf) for additional details, simulations parameters, and initial conditions bounds.
We note that, from the ESC DS dataset, we select only the trajectories characterized by a load factor less then three which ensures that the generated trajectories are within the albatross' physical limits since higher values might result in catastrophic stress on the albatross' wings \cite{richardson2022observations}. 

The data in \cite{richardson2018flight} consist of GPS tracker and activity logger data, gathered during foraging trips of 24 male and 22 female WAN breeding in Bird Island, South
Georgia. Since the dataset includes DS maneuvers in all different direction and phases, we isolate the one traveling upwind and across-wind while facing the wind direction. Further details are provided in the supplemental material (file1.pdf).
Fig. \ref{fig:fig3} (a), displays each dataset samples of airspeed with respect to wind in the direction of flight, computed as in \cite{richardson2018flight}, and the respective regression lines. Although the samples of the scatter plot and histograms pertain to a single uniform sampling in step (iv) of the dataset generation, the regression lines and mean lines are given as the mean of 10 different uniform sampling, so to avoid any bias that might be induced by a single measurement. From Fig. \ref{fig:fig3} it is shown that both ESC systems, with $J_{\dot e}$ and $J_{\dot W}$, capture the overall relationship of groundspeed with respect to wind and, their resulting $V_a$, $V_g$, and $V_{W_{\text{ref}}}$ distributions, closely resemble the distributions of the biological data. We note that the higher accumulation of $V_a$ in Fig. \ref{fig:fig3} (b) for $J_{\dot e}$ and $J_{\dot W}$, is due to the initial condition discretizations of the dataset generation. 
Nevertheless, it is verified that, for both $J_{\dot e}$ and $J_{\dot W}$, the regression line percentage error of the slope and intercept, is lower than 5\%, reflecting a remarkable prediction accuracy for the ESC model. Moreover, the histograms mean values are generally close to the biological data and the $L_2$ distance and Wasserstein distance $D_W$ never exceed the dataset $V_{a_0}$ and $V_{W_{\text{ref}}}$ discretizations value of 2 m/s. Further details are provided in the supplemental material (file1.pdf).
Overall, the $J_{\dot e}$ ESC data provided consistent agreement across all variables while the $J_{\dot W}$ ESC system showed improved agreement for $V_a$ but substantially larger deviation in $V_{W_{\text{ref}}}$.

\begin{figure}[ht]                        
    \centering	\includegraphics[width=0.45\textwidth]{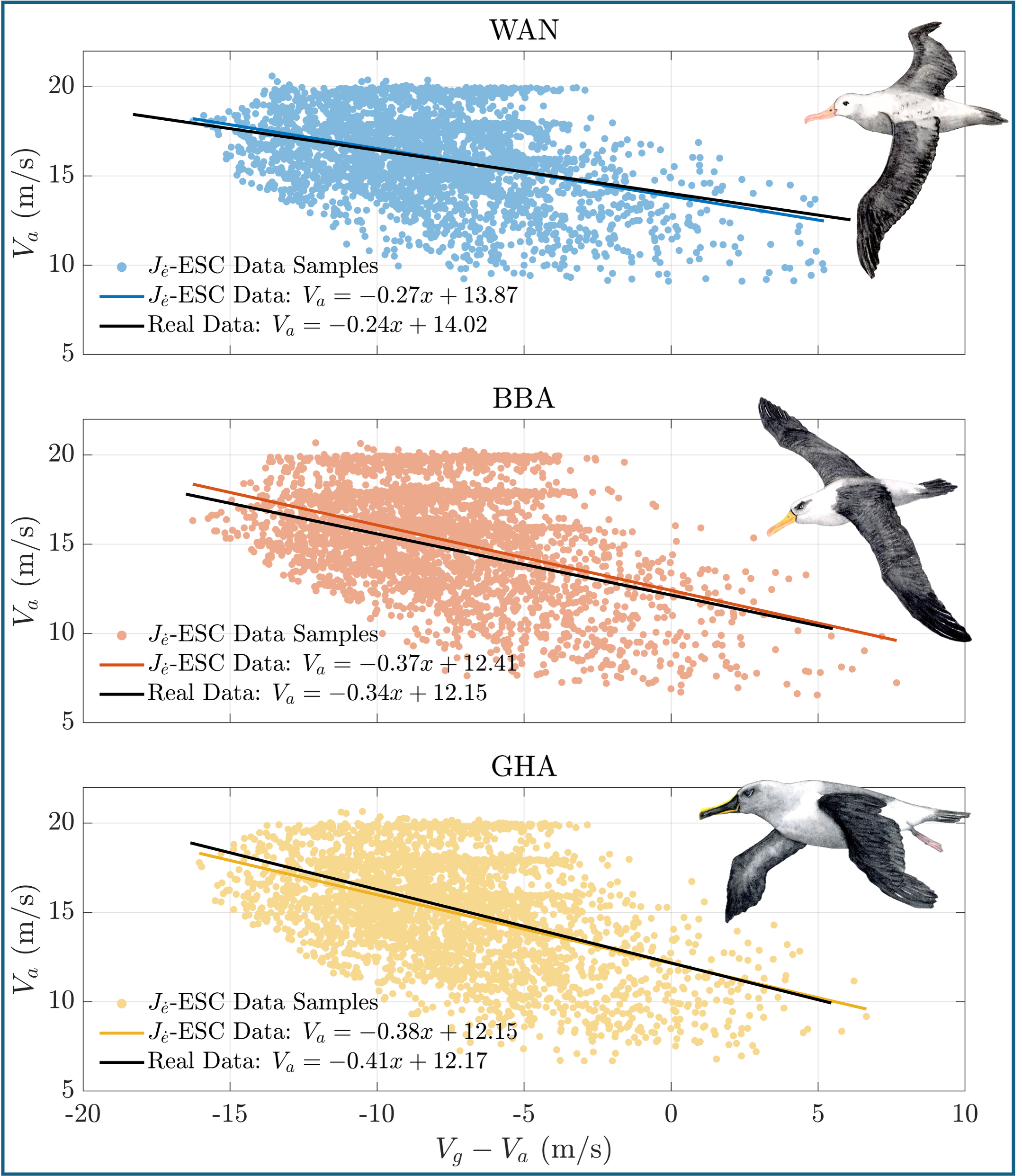}
        {\footnotesize Albatross illustrations credits: Claudia Acutis.}
\caption{\footnotesize{Comparison of ESC DS dataset, using measurements of $J_{\dot e}$, with biological data from \cite{pennycuick1982flight}. Scatter plots of airspeed, $V_a$, with respect to wind (approximated as the difference of groundspeed and airspeed, $V_g - V_a$, as in \cite{pennycuick1982flight}) samples and respective regression lines. From top to bottom, plots refer to the three albatross species WAN, BBA, and GHA, respectively. Percentage errors, in terms of slope and intercept, of less than 10\% highlight that the ESC DS dataset overall captures the DS macroscopic physical relation of airspeed with respect to wind.}}
    \label{fig:4}
\end{figure}

We proceed by comparing the ESC system to the biological data of WAN, BBA, and GHA in \cite{pennycuick1982flight}. Given the number of species, for computational economy, we carry on a comparison solely to the $J_{\dot e}$ ESC system. Since the dataset from \cite{pennycuick1982flight} is not currently available we can extract the regression lines from \cite[Fig. 11]{pennycuick1982flight}. For these three species we select, for our analysis, the maneuvers performed over the sea, which data are gathered from a fixed land site since the author refers to these as the most accurate measurement.
Similar to \cite[Fig. 9]{pennycuick1982flight}, Fig. \ref{fig:4} displays the airspeed with respect to wind for each of the albatross species. Once again, the ESC model is able to capture the overall relationship of airspeed with respect to wind, with regression line slope and intercept error lower than about 10\% and mean airspeed error less than the discretizations size of 2m/s. Additional details on the dataset comparison and specific error values are provided in the supplemental material (file1.pdf).
\begin{remark}\label{rm:5}
    Even with simple point-mass, logarithmic wind profile, and approximate formulas for drag and lift forces, the generated ESC DS dataset successfully match and predict real soaring flight physics on a macroscopic scale, replicating the intrinsic relationship of groundspeed and airspeed with respect to wind conditions.
\end{remark}
\begin{remark}\label{rm:6}
The generated ESC DS system dataset successfully captures the biological characteristics of the DS flight by replicating the behavioral distribution of wind conditions in which the soaring birds choose to fly and the respective speed adjustments.  
\end{remark}
\begin{remark}\label{rm:7}
Considering a conservative 0.02 s computation time of a Simulink ESC simulation, creating the entire dataset requires about 4hrs. On the other end, generating the same dataset using  GPOPs II, would have required around 158 days (considering an average computation time of 20s). The proposed mechanism can generate, predict, and solve DS maneuvers in a way that was never possible using previous literature.
\end{remark}

\textit{Final Remarks}--
To the best of the author's knowledge, this work is the first of its kind in reproducing DS maneuver using a real-time $C_L$ feedback mechanism. 
To address Hypothesis \ref{hp:2}, we showed that the presented ESC formulation is an accurate physical model of real-time evolution and adaptation of $C_L$ which results in optimal energy gain comparable to the non-real-time open loop optimal control counterpart. Moreover, from Remark \ref{rm:2} and \ref{rm:5}, we highlight how the microscopic physical model replicates the macroscopic physics as well. Next, Remark \ref{rm:3} and \ref{rm:6}, and the fact that ESC has recently had numerous successes in replicating biological processes \cite{4dm4-kc4g,abdelgalil2022sea,fish1}, this supports Hypothesis \ref{hp:3}, of the ESC mechanism being rooted in the biology of DS. Lastly, representing the soaring bird dynamics as an ESC system, we \textit{generated a dataset of more than half a million trajectories} with 10,691 net-energy-gaining DS maneuvers. It is clear from Remark \ref{rm:1}, \ref{rm:4}, and \ref{rm:7}, that a study of this kind \textit{would not} be feasible without the proposed ESC framework. Hence, as per Hypothesis \ref{hp:1}, representing optimization processes in DS and beyond as ESC systems can open new doors for computational physics and computational biology.

\bibliographystyle{apsrev4-2}
%

\pagebreak
\pagenumbering{gobble}
\widetext
\begin{center}
\textbf{\large Supplementary Material for ``Universal Extremum Seeking Mechanism for Lift Variation in Soaring Birds Flight: A New Paradigm in Computational Physics and Biology\bm{$^*$}"}

\bigskip
Simone Martini, Dipesh Kunwar, Sameh A. Eisa

\textit{Department of Aerospace Engineering and Engineering Mechanics,\\
University of Cincinnati, Cincinnati, OH 45221 USA}
\end{center}

\section{PC Setup}
Numerical simulations for GPOPS-II and the ESC system are performed in MATLAB and MATLAB Simulink, respectively. Our software configuration uses MATLAB 2019b on Windows 11, while the hardware configuration is: Intel(R) Core(TM) Ultra 9 285K (3.70 GHz), 256GB DDR5 RAM, NVIDIA GeForce RTX 5070 Ti.

\section{Simulations and ESC DS dataset parameters, bounds, and details}

 Since \cite{richardson2018flightP2} and \cite{pennycuick1982flightP2} consider different parameters for WAN, we refer to Table \ref{tab:morphology} WAN (a) and WAN (b), for the WAN species in \cite{pennycuick1982flightP2} and \cite{richardson2018flightP2}, respectively.
 
 Table \ref{tab:params} lists the simulations parameters and initial conditions bounds. The remaining environmental parameters are set as $g = 9.81\,\text{m}\,\text{s}^{-2}$, $\rho = 1.24\, \text{kg}\,\text{m}^{-3}$, and birds aerodynamics parameters are set to $C_{D0} =0.02$ and $e=0.9$. Lastly, for simulations related to the comparison with \cite{richardson2018flightP2}, $z_{\text{ref}}=10$ m and $h_0 = 0.03$ m, while for the analysis related to \cite{pennycuick1982flightP2}, $z_{\text{ref}}=31.5$ m and $h_0 = 10^{-3}$ m. 

For the comparison with \cite{richardson2018flightP2}, in light of computational economy, we isolate the data-points for which $V_{W_{\text{ref}}}<15\, \text{m/s}$, since about $97.5\%$ of the data falls within this range. Considering that the resulting subset will represent both upwind flight and across-wind while facing the wind direction, the initial heading bounds of the ESC dataset, as shown in Table \ref{tab:params}, are selected to include initial heading in the across-wind direction ($0\le\phi_0\le 45, 135\le\phi_0\le 180\,\text{deg}$) besides the upwind one ($135\ge\phi_0\ge45\,\text{deg}$). Moreover to include only simulations traveling and facing upwind, we isolate the samples for which $180>\psi(t)>0$ and $\dot y(t)>0$. 
Table \ref{tab:comparison} reports, for both $J_{\dot e}$ and $J_{\dot W}$, the regression line percentage error of slope $\mathbf{e_{a\%}}$ and intercept $\mathbf{e_{b\%}}$ along with the histograms $L_2$ distance and Wasserstein distance $D_W$ of $V_{a}$, $V_{g}$, and $V_{W_{\text{ref}}}$.

For comparison with \cite{pennycuick1982flightP2}, the author points out that most of the gathered data are for birds flying upwind \cite{pennycuick1982flightP2}. Hence, in generating our dataset, we only consider trajectories, as shown in Table \ref{tab:params}, with initial upwind heading ($135\ge\phi_0\ge45\,\text{deg}$). Moreover from \cite[Fig. 15]{pennycuick1982flightP2}, we can estimate the minimum airspeed of gathered data for each species. Hence, we isolate the samples within those ranges, namely $V_{a_{\text{WAN}}}\ge 8.98$, $V_{a_{\text{BBA}}}\ge 6.48$, $V_{a_{\text{GHA}}}\ge 6.54$. 
Table \ref{tab:jdot_errors} shows the slope and intercept percentage error of the regression lines of each albatross species (WAN, BBA, GHA), resulting from the comparison of the $J_{\dot e}$-ESC dataset and the data from \cite{pennycuick1982flightP2}, and the respective mean airspeed error.

\bibliographystyle{apsrev4-2}

\newpage

\begin{table}[htb]
\centering
\caption{Morphological parameters from Paper~1 and Paper~2.}
\label{tab:morphology}
\begin{tabular}{lcccc}
\hline
 & \multicolumn{3}{c}{Comparison with \cite{pennycuick1982flightP2}} & Comparison with \cite{richardson2018flightP2} \\
\cline{2-4} \cline{5-5}
 & WAN (a) & BBA & GHA & WAN (b) \\
\hline
$m$ (kg)  & 8.73 & 3.79 & 3.79 & 12 \\
$S$ ($\text{m}^2$)  & 0.611 & 0.356 & 0.352 & 0.84
 \\
$AR$ & 15.0 & 13.1 & 13.5 & 12.56
 \\
\hline
\end{tabular}
\end{table}
\begin{table}[htb]
\centering
\caption{Simulation parameters and discretization for Paper~1 and Paper~2.}
\label{tab:params}
\begin{tabular}{lcccccc}
\toprule
& \multicolumn{3}{c}{Comparison with \cite{pennycuick1982flightP2}} & \multicolumn{3}{c}{Comparison with \cite{richardson2018flightP2}} \\
\cmidrule(lr){2-4} \cmidrule(lr){5-7}
\textbf{} & min & max & $d$ & min & max & $d$ \\
\midrule

$V_{W_{\text{ref}}}$ (m/s) 
& 2 & 16 & 2 
& 3 & 17 & 2 \\

$z_0$ (m/s) 
& 0.5 & 8 & 0.5 
& 0.5 & 8 & 0.5 \\

$V_0$ (m/s) 
& 12 & 20 & 2 
& 12 & 20 & 2 \\

$\phi_0$ (deg) 
& -45 & 45 & 11.25 
& -45 & 45 & 11.25 \\

$\psi_0$ (deg) 
& 45 & 135 & 22.5 
& 0 & 180 & 22.5 \\

$C_{L_0}$ 
& 0.5 & 1 & 0.25 
& 0.5 & 1 & 0.25 \\

\midrule
\textbf{Total combinations} 
& \multicolumn{3}{c}{$103,680$} 
& \multicolumn{3}{c}{$186,624$} \\

\midrule
\textbf{Total Simulations} 
& \multicolumn{3}{c}{$311,040$} 
& \multicolumn{3}{c}{$373,248$} \\

\bottomrule
\end{tabular}
\end{table}

\begin{table}[htb]
\centering
\caption{Regression line errors and histogram metrics for ESC datasets with respect to \cite{richardson2018flightP2}.}
\label{tab:comparison}
\begin{tabular}{l|c|c}
\hline
\textbf{Dataset} & \textbf{Regression} & \textbf{Histograms} \\
\hline
$J_{\dot e}$ &
\begin{tabular}{c c}
$\mathbf{e_{a\%}}$ & $\mathbf{e_{b\%}}$ \\
4.83.45 & 2.36 
\end{tabular} &
\begin{tabular}{c c c}
\textbf{Variable} & $L_2$ & $D_W$ \\
$V_{Wref}$ & 0.111 & 0.512 \\
$V_g$      & 0.166 & 1.007 \\
$V_a$      & 0.203 & 1.209
\end{tabular} \\
\hline
$J_{\dot W}$ &
\begin{tabular}{c c}
$\mathbf{e_{a\%}}$ & $\mathbf{e_{b\%}}$ \\
-1.60 & -3.32 
\end{tabular} & 
\begin{tabular}{c c c}
\textbf{Variable} & $L_2$ & $D_W$ \\
$V_{Wref}$ & 0.278 & 1.955 \\
$V_g$      & 0.162 & 1.123 \\
$V_a$      & 0.070 & 0.452
\end{tabular} \\
\hline
\end{tabular}
\end{table}

\begin{table}[htb!]
\centering
\caption{Regression line errors and mean airspeed error with respect data from \cite{pennycuick1982flightP2}.}
\label{tab:jdot_errors}
\begin{tabular}{l c c c}
\hline
 & $\mathbf{e_{a\%}}$ & $\mathbf{e_{b\%}}$ & $\mathbf{e_{V_a,\textbf{avg}}}$ \\
\hline
WAN & 10.31 & -1.09 & 1.06 \\
BBA & 6.76  & 2.14  & 1.58 \\
GHA & -6.61 &-0.20  & 1.22 \\
\hline
\end{tabular}
\end{table}
\end{document}